\documentclass[aps,prb,superscriptaddress,twocolumn,showpacs]{revtex4}

\usepackage{bm}
\usepackage{graphicx}

\begin{document}

\title{Anisotropy and Ising-like transition of the $S=5/2$
    two-dimensional Heisenberg antiferromagnet Mn-formate di-Urea}
\author{Alessandro Cuccoli}
\affiliation{Dipartimento di Fisica dell'Universit\`a di Firenze,
    Via G. Sansone 1, I-50019 Sesto Fiorentino (FI), Italy}
\affiliation{Istituto Nazionale per la Fisica della Materia (INFM),
  Unit\`a di Ricerca di Firenze, Via G. Sansone 1, I-50019 Sesto
  Fiorentino (FI), Italy}

\author{Tommaso Roscilde}
\affiliation{Dipartimento di Fisica "A. Volta" dell'Universit\`a di Pavia,
    via A. Bassi 6, I-27100 Pavia}
\affiliation{Istituto Nazionale per la Fisica della Materia (INFM),
  Unit\`a di Ricerca di Pavia, via A. Bassi 6, I-27100 Pavia}

\author{Valerio Tognetti}
\affiliation{Dipartimento di Fisica dell'Universit\`a di Firenze,
    Via G. Sansone 1, I-50019 Sesto Fiorentino (FI), Italy}
\affiliation{Istituto Nazionale per la Fisica della Materia (INFM),
  Unit\`a di Ricerca di Firenze, Via G. Sansone 1, I-50019 Sesto
  Fiorentino (FI), Italy}

\author{Ruggero Vaia}
\affiliation{Istituto di Fisica Applicata `N.~Carrara'
             del Consiglio Nazionale delle Ricerche,
             via Panciatichi~56/30, I-50127 Firenze, Italy}
\affiliation{Istituto Nazionale per la Fisica della Materia (INFM),
  Unit\`a di Ricerca di Firenze, Via G. Sansone 1, I-50019 Sesto
  Fiorentino (FI), Italy}

\author{Paola Verrucchi}
\affiliation{Dipartimento di Fisica dell'Universit\`a di Firenze,
    Via G. Sansone 1, I-50019 Sesto Fiorentino (FI), Italy}
\affiliation{Istituto Nazionale per la Fisica della Materia (INFM),
  Unit\`a di Ricerca di Firenze, Via G. Sansone 1, I-50019 Sesto
  Fiorentino (FI), Italy}

\date{\today}

\begin{abstract}
Recently reported measurements of specific heat on the compound
Mn-formate di-Urea (Mn-f-2U) by Takeda {\em et al.} [Phys. Rev. B {\bf
63}, 024425 (2001)] are considered. As a model to describe the overall
thermodynamic behavior of such compound, the easy-axis two-dimensional
Heisenberg antiferromagnet is proposed and studied by means of the {\em
pure quantum self-consistent harmonic approximation} (PQSCHA). In
particular it is shown that, when the temperature decreases, the
compound exhibits a crossover from 2D-Heisenberg to 2D-Ising behavior,
followed by a 2D-Ising-like phase transition, whose location allows to
get a reliable estimate of the easy-axis anisotropy driving the
transition itself. Below the critical temperature
$T_{_{\rm{N}}}=3.77$~K, the specific heat is well described by the
two-dimensional easy-axis model down to a temperature $T^*=1.47$~K
where a $T^3$-law sets in, possibly marking a low-temperature crossover
of magnetic fluctuations from two to three dimensions.
\end{abstract}


\maketitle

The quantum Heisenberg antiferromagnet (QHAF) on the square lattice is
one of the most widely investigated magnetic models because of both its
fundamental theoretical properties and the existence of many real
compounds whose magnetic behavior is ruled by a spin-spin interaction
properly described by the 2D QHAF Hamiltonian. Due to their layered
structures, such compounds have an intralayer exchange integral $J$
which is much larger than the interlayer one, $J'$. This ensures a 2D
thermodynamic behavior to persist down to a certain crossover
temperature which could be naively estimated of the order of $J'S^2$.
However, most of the above mentioned compounds display a phase
transition towards an ordered phase at a critical temperature of the
order of $JS^2$.

If one assumes a fully isotropic intralayer interaction, which cannot
induce any finite temperature transition by itself, this experimental
finding may be explained by noticing that strong correlations between
spins on a single layer act as an effective amplifier of the interlayer
interaction, in that the coupling between neighboring spins on
different layers drags in a similar coupling the surrounding spins
within a distance of the order of the 2D magnetic correlation
length~\cite{VillainL77}. However, most of the experimentally observed
phase transitions do also display features which suggest a persistent
2D behavior, and are not compatible with the above sketched mechanism.
On the other hand, such experimental observations may be explained by
allowing the intralayer interaction to be anisotropic; the existence of
some anisotropic coupling may hence be invoked, and a more detailed
analysis developed.

In this paper we consider the quasi-two-dimensional real compound
Mn(HCOO)$_2\cdot$2(NH$_2$)$_2$CO (Mn-formate di-Urea, or Mn-f-2U),
which is is a remarkable realization of a $S=5/2$ 2D QHAF in a wide
temperature region. From specific-heat and susceptibility
measurements~\cite{TakedaEtal89,TakedaEtal00,TakedaEtal01} the
intralayer spin-spin coupling $J$ turns out to be $J=0.68\pm{0.04}$~K,
while $J'$ is orders of magnitude lower. Characteristic features of a
phase-transition, such as a sharp peak in the specific heat, are also
observed at $T_{_{\rm{N}}}=3.77\pm{0.02}$~K and
suggested~\cite{TakedaEtal00} the existence of an easy-axis anisotropy
in the intralayer spin-spin coupling. By proton NMR measurements it has
been established in Ref.~\onlinecite{KuboEtal00} that the interlayer
coupling is much smaller than the anisotropy, which has been estimated
to be of the order of $10^{-2}J$.

We hence propose, as a model for the thermodynamics of Mn-f-2U, the
$S=5/2$ QHAF on the square lattice with easy-axis exchange anisotropy
(EA-QHAF), whose Hamiltonian reads
\begin{equation}
 \hat{\cal H}={J\over 2}\sum_{{\bm i},{\bm d}}\Big[\mu\big(
 \hat{S}_{\bm i}^x\hat{S}_{{\bm i}+{\bm d}}^x+
 \hat{S}_{\bm i}^y\hat{S}_{{\bm i}+{\bm d}}^y\big)+
 \hat{S}_{\bm i}^z\hat{S}_{{\bm i}+{\bm d}}^z\Big]~,
 \label{e.EAQHAF}
\end{equation}
where ${\bm{i}}=(i_1,i_2)$ runs over the sites of a square lattice,
${\bm{d}}$ connects each site to its four nearest neighbors, $J>0$ is
the antiferromagnetic exchange integral and $\mu\in[0,1)$ is the
easy-axis anisotropy parameter. The spin operators
$\hat{S}^\alpha_{\bm{i}}$ ($\alpha=x,y,z$) are such that
$|\hat{\bm{S}}|^2=S(S+1)$ and obey
$[\hat{S}^\alpha_{\bm{i}},\hat{S}^\beta_{\bm{j}}]=i\,
\varepsilon_{\alpha\beta\gamma}\delta_{\bm{ij}}\hat{S}^\gamma_{\bm{i}}$.

Classical~\cite{MC_EA} and quantum~\cite{QMC_EA} Monte Carlo
simulations predict such model to display a 2D-Ising phase transition
at a finite temperature $T_{_{\rm{I}}}(S,\mu)$, which continuously
decreases as $\mu$ increases, finally vanishing for $\mu\to1$, i.e., in
the isotropic model. The picture in the case of Mn-f-2U is hence that
its 2D-Ising transition at $T_{_{\rm{I}}}$ immediately triggers the
observed~\cite{TakedaEtal89,TakedaEtal00} phase transition to 3D
long-range order: therefore, we assume $T_{_{\rm{I}}}=T_{_{\rm{N}}}$.

We use the {\em pure-quantum self-consistent harmonic approximation}
(PQSCHA)~\cite{PQSCHA}, a semiclassical method which reduces the
expressions of quantum statistical averages to effective classical-like
ones, where temperature- and spin-dependent renormalization parameters
appear; the thermodynamics of the effective model can then be studied
by means of classical techniques, like classical Monte Carlo
simulations. Besides a uniform additive term that does not affect the
evaluation of statistical averages, the PQSCHA effective Hamiltonian
for the EA-QHAF reads~\cite{CRTVV01}
\begin{equation}
 {\cal H}_{\rm eff}=
 -\frac{J_{\rm eff}\widetilde{S}^2}2 \sum_{{\bm i},{\bm d}}\Big[
 \mu_{\rm eff}\big(s_{\bm i}^xs_{{\bm i}+{\bm d}}^x{+}
 s_{\bm i}^ys_{{\bm i}+{\bm d}}^y\big)+
 s_{\bm i}^zs_{{\bm i}+{\bm d}}^z\Big]~,
\label{e.Heff1}
\end{equation}
where $\widetilde{S}=S+1/2$ and
${\bm{s}}_{\bm{i}}=(s^x_{\bm{i}},s^y_{\bm{i}},s^z_{\bm{i}})$ are
classical unit vectors. $J_{\rm{eff}}(t,S,\mu)<J$ and
$\mu_{\rm{eff}}(t,S,\mu)>\mu$ are the renormalized exchange and
anisotropy of the effective classical model, respectively. Hereafter
the dimensionless temperature $t=T/J\widetilde{S}^2$ is used.

This method was applied to the EA-QHAF, and, in particular, to the
study of the $\mu$- and $S$-dependence of the Ising critical
temperature, which is shown in the phase diagram $t_{_{\rm{I}}}(S,\mu)$
reported in Ref.~\onlinecite{CRTVV01}. Using that phase diagram, the
anisotropy value corresponding to the experimental critical temperature
of Mn-f-2U, $t_{_{\rm{I}}}=0.616\pm0.039$, is
$\mu=0.981^{+0.014}_{-0.029}$, the large error being due to the
experimental uncertainty on $J$. This anisotropy estimate is close to
that of the compound Rb$_2$MnF$_4$, a $S=5/2$ EA-QHAF we have
extensively investigated in Ref.~\onlinecite{CRTVV00}, namely
$\mu=0.9942$. Moreover, performing a new set of classical MC
simulations with a slightly different value of $\mu$, more fit to the
case of Mn-f-2U, is, at present, too large an effort compared to the
expected benefits. As a matter of fact, due to the poor knowledge of
$J$, any choice of the value of $\mu$ in the interval estimated above
is affected by a high degree of arbitrariness. We will hence hereafter
use the already known PQSCHA curves~\cite{CRTVV00} for $\mu=0.9942$;
owing to the proximity between the two anisotropies (i.e.,
$\mu=0.981^{+0.014}_{-0.029}$ estimated for Mn-f-2U from our phase
diagram~\cite{CRTVV01}, and $\mu=0.9942$ corresponding to our previous
theoretical data~\cite{CRTVV00}), we expect to be able to single out
the characteristic features of the experimental data by comparison with
the theoretical curves.

\begin{figure}
\includegraphics[bbllx=11mm,bblly=19mm,bburx=179mm,bbury=152mm,%
     width=86mm,angle=0]{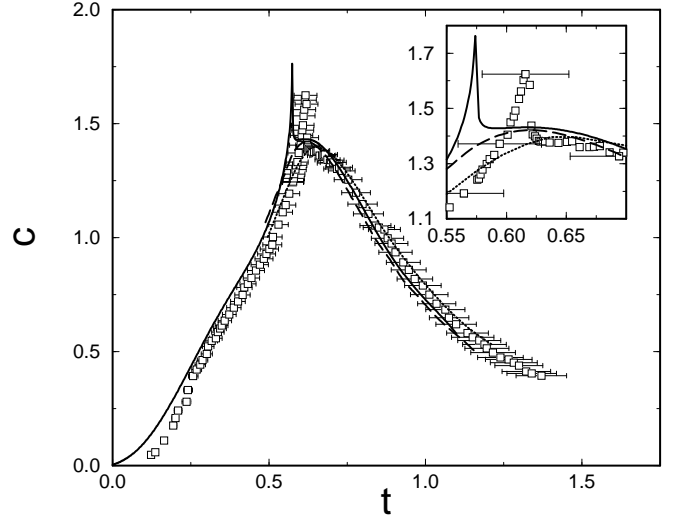}
\caption{\label{f.norescaled}
Specific heat {vs} $t=T/J\widetilde{S}^2$, for $S$=5/2. Mn-f-2U
experiments (squares)~\cite{TakedaEtal01}, EA-QHAF with $\mu=0.9942$
(solid line)~\cite{CRTVV01}, isotropic QHAF (dashed line), isotropic
QHAF with improved PQSCHA (dotted line)~\cite{CTVV97}. Error bars are
due to the experimental uncertainty on $J$ for Mn-f-2U. Inset: zoom of
the Heisenberg-Ising crossover region, with few representative error
bars.}
\end{figure}

Focusing our attention on recent measurements~\cite{TakedaEtal01} of
the magnetic specific heat of Mn-f-2U, we make a twofold comparison of
such data with our predictions for the EA-QHAF. The first comparison,
shown in Fig.~\ref{f.norescaled}, is made between the measured specific
heat of Mn-f-2U and the theoretical one with $\mu=0.9942$ as a function
of the dimensionless temperature. Besides these data, two PQSCHA curves
for the $S=5/2$ isotropic model are also reported: the dashed one is
obtained within the same approximation level as for the anisotropic
model, while the dotted one is obtained by an improved version of the
PQSCHA scheme~\cite{CTVV97} that was only applied to the isotropic
case; the latter was already compared to the experimental data by
Takeda {\em et al.}~\cite{TakedaEtal01}, so it is included in order to
avoid mistakes. Both the theoretical specific heat for the EA-QHAF with
$\mu=0.9942$ and the measured specific heat of Mn-f-2U are seen to
behave as that of the isotropic QHAF for temperatures well above
$t_{_{\rm{I}}}$. The agreement with the isotropic behavior, however,
gets worse once the anisotropy starts to drive the anisotropic systems
towards the Ising critical regime. The temperature where the
anisotropic (solid) curve deviates from the isotropic (dashed) one to
form the Ising spike marks a crossover between the Heisenberg and the
Ising regime. This crossover, the position of the peak, and the whole
behavior below the critical temperature depend on the anisotropy value
$\mu$. We stress that if we had calculated the PQSCHA curve for the
value $\mu=0.981$ estimated from the PQSCHA phase
diagram~\cite{CRTVV01}, the position of the theoretical peak would
coincide with that of the experimental one.

\begin{figure}
\includegraphics[bbllx=9mm,bblly=11mm,bburx=181mm,bbury=152mm,%
 width=86mm,angle=0]{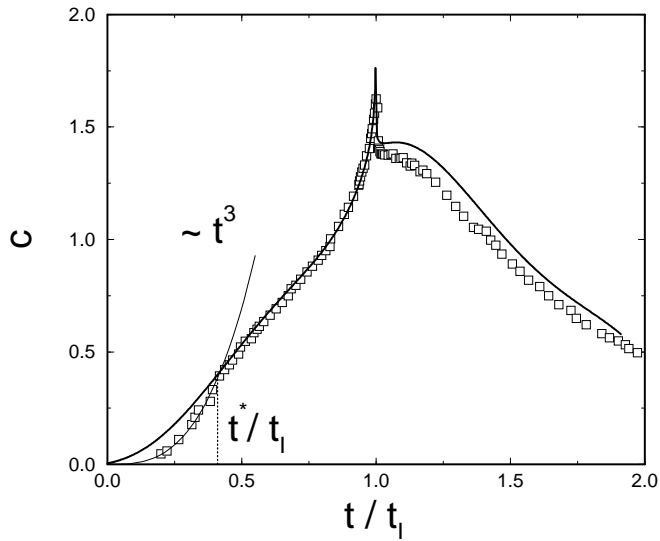}
\caption{\label{f.rescaled}
Specific heat {\em vs} $t/t_{_{\rm{I}}}$, with
$t_{_{\rm{I}}}\equiv{T}_{_{\rm{I}}}/J\widetilde{S}^2=0.616$ for the
experimental data, and $t_{_{\rm{I}}}=0.575$ for the theoretical line.
The dashed line is a best fit of the first 8 experimental data points
to a $t^3$ law; other symbols are as in Fig.~\ref{f.norescaled}.}
\end{figure}

In order to emphasize the common features of the theoretical and the
experimental curves in the Ising regime we scale their abscissa $t$
with the actual critical temperatures, i.e.,
$t_{_{\rm{I}}}\equiv{T}_{_{\rm{I}}}/J\widetilde{S}^2=0.616$ for the
experimental data, and $t_{_{\rm{I}}}=0.575$ for the theoretical ones,
making the peaks match as shown in Fig.~\ref{f.rescaled}. In this way,
remarkable agreement between experiment and theory shows up not only
for the shape of the transition peak but also for the whole temperature
region between $t/t_{_{\rm{I}}}=1$, and
$t^*/t_{_{\rm{I}}}\simeq{0.42}$, where the experimental data for the
compound Mn-f-2U depart from the PQSCHA curve.

These results show that the compound Mn-f-2U, besides behaving as a 2D
EA-QHAF above and in the critical region, displays such behavior in its
specific heath also below $t_{_{\rm{I}}}$, down to a much lower
temperature $t^*\simeq0.24$. This suggests that below $t_{_{\rm{I}}}$,
even though the weak interlayer coupling drives the system into a 3D
ordered phase, magnetic fluctuations, upon which the specific heat
directly depends, remain confined within the layers, being those
between different layers incoherent. The 2D picture of the ordered
phase eventually breaks down at $t=t^*$. Below this point the
experimental data for Mn-f-2U are much better described by a simple
$t^3$-law; such law characterizes the low-$t$ region of a 3D quantum
antiferromagnet, with the main contribution arising from linear
excitations. The temperature $t^*$ could thus be interpreted as marking
the dimensional crossover from 2D to 3D behavior, corresponding to the
onset of coherence in the magnetic excitations propagating
perpendicularly to the layers. However, at this stage, the microscopic
mechanism responsible for the observed low-$t$ behavior cannot be
firmly identified; nonetheless, the above picture seems to be the
simplest one that accounts for the experimentally observed behavior.

In conclusion, we have compared the theoretical findings based on the
PQSCHA for the 2D EA-QHAF with recent specific heat data on Mn-f-2U.
The comparison reveals the existence of a crossover from a
high-temperature 2D-Heisenberg regime to a critical 2D-Ising regime
that triggers the phase transition at $T_{_{\rm{N}}}=3.77$~K observed
in Ref.~\onlinecite{TakedaEtal89}. In the ordered phase, the specific
heath still behaves as that of a 2D EA-QHAF down to the temperature
$T^*=J\widetilde{S}^2\,t^*\simeq1.47$~K, where 3D behavior sets in. The
success of the EA-QHAF model in describing the specific heat of Mn-f-2U
in a wide temperature range makes trustworthy our prediction for the
value of the anisotropy parameter, $\mu=0.981$, as obtained above by
the sole knowledge of $t_{_{\rm{I}}}=T_{_{\rm{I}}}/(J\widetilde{S}^2)$.
The above value of $\mu$ implies the anisotropy to be about $1.9$\% of
the exchange energy. Accounting for the large uncertainty upon the
estimated $J$, however, the actual anisotropy could range from $0.6$\%
to $4.5$\%. Anyhow, the above estimate of the anisotropy strength is
compatible with the one made by Kubo {\em et~al.} in
Ref.~\onlinecite{KuboEtal00}. Experiments on Mn-f-2U aimed at a more
precise determination of both $J$ and $\mu$, as well as at the
characterization of magnetic excitations, such as inelastic neutron
scattering ones, come now to be highly desirable.

\begin{acknowledgments}
We thank prof. K. Takeda for sending us the experimental data on
Mn-f-2U, published in Ref.~\onlinecite{TakedaEtal01}.  This work has
been partially supported by the COFIN2000-MURST fund, and by the CRUI
within the VIGONI programme.
\end{acknowledgments}

\end{document}